\documentstyle[preprint,aps,epsfig,epsf,rotate]{revtex}

\newcommand{\sfrac}[2]{\mbox{\footnotesize $\frac{#1}{#2}$}}

\begin{document}
\tighten

\preprint{
\vbox{ 
%\hbox{\today}
\hbox{ADP-01-31/T463}
\hbox{JLAB-THY-01-22}
}}

% \wideabs{

\title{Parton Distributions from Lattice QCD}

\author{W.~Detmold$^1$, W.~Melnitchouk$^{1,2}$ and A.W.~Thomas$^1$}
\address{$^1$ Special Research Centre for the Subatomic Structure of
        Matter, and Department of Physics and Mathematical Physics,
        Adelaide University, 5005, Australia}
\address{$^2$ Jefferson Lab, 12000 Jefferson Avenue,
        Newport News, VA 23606, USA}

\maketitle

\begin{abstract}
We extract the $x$ dependence of the valence nonsinglet $u-d$ distribution
function in the nucleon from the lowest few moments calculated on the
lattice, using an extrapolation formula which ensures the correct
behavior in the chiral and heavy quark limits.
We discuss the implications for the quark mass dependence of meson masses
lying on the $J^{PC}=1^{--}$ Regge trajectory.
\end{abstract}

\newpage
%%%%%%%%%%%%%%%%%%%%%%%%%%%%%%%%%%%%%%%%%%%%%%%%%%%%%%%%%%%%%%%%%%%%%%%%%
\section{Introduction}

Parton distribution functions (PDFs) contain a wealth of information
on the nonperturbative structure of the nucleon.
Quark and gluon distributions probed in deep inelastic scattering and
other high energy processes have provided valuable insights into the 
workings of QCD in the low energy domain.
The observation of an asymmetry between $\bar d$ and $\bar u$ quarks in 
the proton sea \cite{NMC,E866}, to take just one example, has served to 
highlight the important role that the pion cloud of the nucleon and the 
chiral symmetry of QCD \cite{TMS} plays in hadronic structure, even at 
high energies.

More generally, studies of PDFs can help with the task of identifying
the appropriate effective degrees of freedom of QCD at low energies. 
Through the application of the operator product expansion (OPE) to QCD,
high energy processes such as deep inelastic scattering can be factorized
into short and long distance contributions, allowing one to calculate the
former in perturbation theory, while isolating all of the nonperturbative
physics in the latter.
Over the past two decades considerable experience has been accumulated 
with various nonperturbative, low energy models of the nucleon which have 
been used to study PDFs \cite{BOOK}.
Initial studies focused on the valence quark distributions as a means
of constraining valence quark model parameters, although recently more
ambitious efforts have attempted to describe sea quark and gluon
distributions from low energy models.

Although the model studies have been helpful in exploring the relationship
between high energy processes and low energy phenomenology, ultimately one
would like a more exact connection of PDFs with QCD.
A mathematically more rigorous approach is provided through lattice QCD.
Indeed, the determination of the moments of the PDFs is one of the
benchmark calculations of hadron structure in lattice QCD.
Modern computational advances have allowed large scale simulations to be
undertaken which are progressively improving the errors associated with 
finite lattice spacings and finite volume effects.
However, until recently \cite{DMNRT} large differences between lattice 
results and experiment have remained.

Because PDFs are light cone correlation functions, it is not possible to 
calculate them directly on the lattice in Euclidean space.
Instead one calculates moments of PDFs, defined (for Bj\"orken $x$) as:
\begin{eqnarray}
    \label{moments}
\langle x^n \rangle_q
&=& \int_0^1 dx\ x^n\ \left( q(x) + (-1)^{n+1} \bar q(x) \right)\ ,
\end{eqnarray}
which are related through the OPE to matrix elements of local
twist-two operators.
A number of calculations of PDF moments have been performed over the last
decade, most notably by the QCDSF group \cite{QCDSF} in the quenched 
approximation.
More recently, the MIT group \cite{MIT} has confirmed the earlier quenched 
results, and in addition made the first unquenched simulations.
These results indicate that at the relatively large quark masses at which 
the calculations were made, the unquenched results are indistinguishable
from the quenched within the current errors.

Despite the impressive progress of lattice calculations of moments of
PDFs, there has been a long standing problem in reconciling the lattice
data with experiment, which has posed a serious threat to the credibility 
of current lattice calculations.
Namely, for the unpolarized moments all of the calculations to date, which 
have been made at quark masses of between 30 and 190~MeV, have yielded
results which are typically 50\% larger than the experimental values when 
linearly extrapolated to the physical quark masses.
This discrepancy was recently resolved with the observation \cite{DMNRT}
that a linear extrapolation in quark mass omits crucial physics associated
with the long range structure of the nucleon in the form of the pion
cloud.
In particular, the inclusion of the nonlinear, nonanalytic dependence on
the quark mass in the extrapolation of the moments from the region of
large quark masses to the physical, light quark masses allows one to fit
both the lattice data and the experimental values with good accuracy
\cite{DMNRT}.
It therefore restores confidence in the extraction of all hadron
observables from current simulations.

In this paper we go one step further and ask what can the lattice data
tell us about the $x$ dependence of the PDFs themselves?
In particular, we shall extract the valence nonsinglet distribution
$u_v-d_v$ by extrapolating the lattice data to the physical region using 
a form that ensures the correct behavior in the chiral and heavy quark 
limits.
In Section~II we examine the initial question of the extent to which the
$x$ dependence can be determined from just a few low moments of a PDF.
After establishing the accuracy to which the PDF can be reconstructed,
we then extract the valence $u_v-d_v$ distribution from the
calculated lattice moments in Section~III.
Using the chiral extrapolation formula, we compute the $x$ dependence of
the $u_v-d_v$ distribution as a function of quark mass.
In Section~IV we discuss the implications of the results for masses of
mesons lying on the $J^{PC}=1^{--}$ Regge trajectory.
Finally, in Section~V we make some concluding remarks.

%%%%%%%%%%%%%%%%%%%%%%%%%%%%%%%%%%%%%%%%%%%%%%%%%%%%%%%%%%%%%%%%%%%%%%%%%
\section{Reconstruction of Distributions from Moments}

The reconstruction of the complete $x$ dependence of a PDF in principle
requires infinitely many moments, so that {\em a priori} the task of 
building up a PDF from only a few, low moments may seem rather ambitious.  
This is especially true for PDFs at large $x$, information on which is 
contained in higher moments.
On the lattice, because of problems with operator mixing for operators
with spin $\geq 5$, all lattice calculations have so far been restricted to
$n \leq 3$.
It is crucial, therefore, if one is to ever make use of lattice data to 
learn about PDFs, to see which features of the PDFs can be reconstructed 
from just the lowest 3 or 4 moments.

Formally, the $x$ dependence of the PDF can be obtained by performing
an inverse Mellin transform of the moments.
If enough moments are known (and can be continued to complex $n$),
one can make a reliable reconstruction of the distributions numerically
(see e.g. Ref.\cite{WM}).
However, since lattice data available at present (and in the foreseeable 
future) only exist for the first three nontrivial moments, this procedure
becomes impractical.
Alternatively, one can reconstruct the parameters in a given PDF
parameterization by working with the moments directly.  
A similar method was applied to the reconstruction of polarized parton
distributions in Ref.\cite{POLRECON}, while in Ref.\cite{WEIGL} the $x$
dependence of PDFs was investigated subject to constraints on their
small- and large-$x$ behavior.
In the present work, however, we wish to determine what constraints on
the $x$ dependence of the valence distribution are imposed by the actual
lattice data.

A typical parameterization of quark (or antiquark) distributions used
in global fits \cite{MRS,CTEQ,GRV} is of the form:
\begin{equation}
  \label{param}
  x q(x) = a x^b (1-x)^c (1 + \epsilon\sqrt{x} + \gamma x)\ ,
\end{equation}
for which the corresponding moments are given by:
\begin{equation}
  \label{momentfit}
  \langle x^n \rangle_q
  = a \left[ B(1+c,b+n)
           + \epsilon B(1+c,\sfrac{1}{2}+b+n)
           + \gamma B(1+c,1+b+n)
      \right]\ ,
\end{equation}
where $B(x,y)$ is the $\beta$-function.
For a given set of PDF moments, the parameters $a, b, c, \epsilon$ and
$\gamma$ can be fitted using Eq.~(\ref{momentfit}), and the corresponding
$x$ distribution obtained from Eq.(\ref{param}).
To test the effectiveness of this method, we calculate moments of PDFs from 
the recent MRS \cite{MRS}, CTEQ \cite{CTEQ} and GRV \cite{GRV} 
parameterizations, and examine how well the reconstructed distributions 
agree with the original parameterizations.

In order to make best use of currently available lattice data, we
concentrate on the nonsinglet valence distribution $u_v(x)-d_v(x)
\equiv [u(x)-\bar u(x)] - [d(x)-\bar d(x)]$.
While the present analysis can be easily extended to the singlet sector, 
the singlet quark distribution on the lattice receives contributions from
disconnected diagrams, corresponding to operator insertions in quark lines 
which are disconnected (except through gluon lines) from the nucleon 
source.
Disconnected diagrams are considerably more difficult to compute, so only
exploratory studies of these have so far been completed \cite{DISCON}.
On the other hand, since the disconnected contributions are flavor
independent (for equal $u$ and $d$ quark masses), they cancel exactly in
the nonsinglet $u-d$ combination.

\begin{figure}
\begin{center}
  \epsfysize=9truecm \leavevmode
  \epsfbox{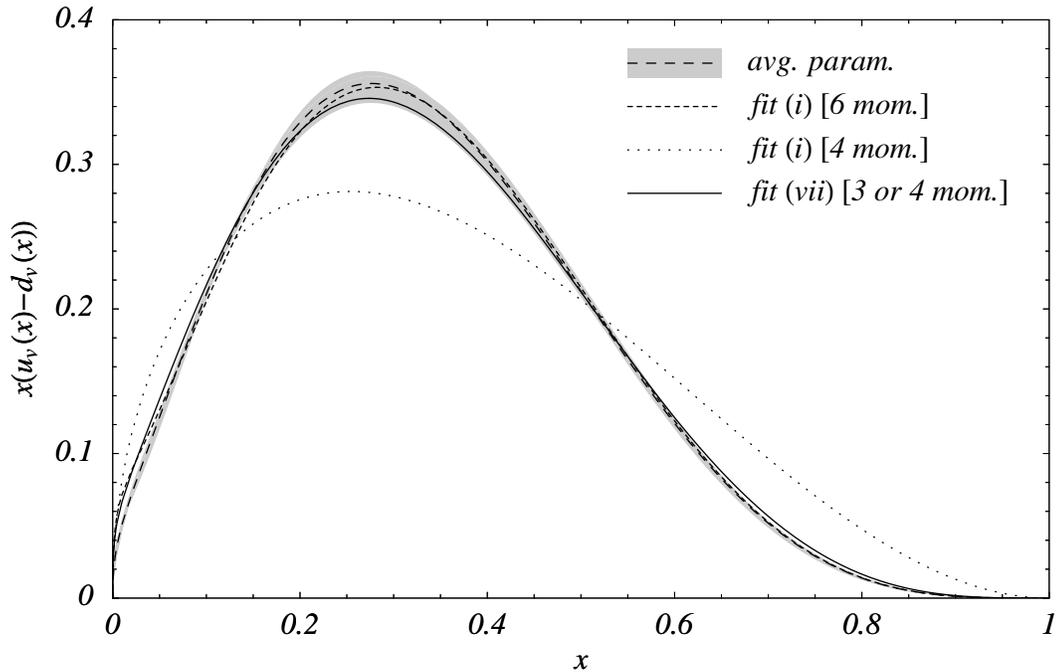}\vspace*{1cm}
\caption{Quality of reconstruction of the valence $x (u_v(x)-d_v(x))$
        distribution from several low moments: the shaded region
        represents the spread between different next-to-leading order
        distributions from global parameterizations
        \protect\cite{MRS,CTEQ,GRV} (at $Q^2=4$~GeV$^2$ in the
        $\overline{\rm MS}$ scheme), while the long--dashed line
        represents a parameterization of the average of the three
        distributions, Eq.~(\protect\ref{avgdiff}).
        The short--dashed line (which is almost indistinguishable from
        the long--dashed, average parameterization) is the distribution
        reconstructed from the lowest six moments of the average
        parameterization using Eq.~(\protect\ref{momentfit}) with
        $\epsilon$ and $\gamma$ unconstrained.
        The dotted curve indicates the fit obtained when only four
        moments are used with the same fitting form.
        In contrast, the solid lines represent the distribution
        reconstructed from the lowest three moments ($n=0, 1, 2$) using
        Eq.~(\protect\ref{momentfit}) with $\epsilon$ and $\gamma$
        constrained to the values obtained from direct fits to the
        average distribution, $\epsilon=\epsilon_\Delta$ and
        $\gamma=\gamma_\Delta$.}
\label{reconstruct}
\end{center}
\end{figure}

To enable comparison with the reconstructed distributions, we employ
Eq.~(\ref{param}) to construct an `average parameterization' from the 
global PDF analyses.
Note that the PDF fits in Refs.\cite{MRS,CTEQ,GRV} are given for $u_v(x)$ 
and $d_v(x)$ separately, whereas here we require a parameterization of the 
difference.
Using an equally weighted numerical fit to the $u_v(x) - d_v(x)$ difference
(at next-to-leading order in the $\overline{\rm MS}$ scheme, at a scale
$Q^2 = 4$~GeV$^2$) from the parameterizations of Ref.\cite{MRS,CTEQ,GRV},
we find an average fit:
\begin{equation}
  \label{avgdiff}
  x(u_v(x)-d_v(x))_{\rm avg}\
  =\ a_\Delta\ x^{b_\Delta} (1-x)^{c_\Delta}
    (1 + \epsilon_\Delta\sqrt{x} + \gamma_\Delta x)\ ,
\end{equation}
with $b_{\Delta}=0.476$, $c_{\Delta}=3.750$,
$\epsilon_{\Delta}=-3.152$, $\gamma_{\Delta}=16.64$, and with
$a_{\Delta}=0.561$ determined by the normalization condition,
$\int_0^1 dx \left(u_v(x)-d_v(x)\right)=1$.
Figure~\ref{reconstruct} shows the average distribution (long-dashed 
line), with the spread between the three parameterizations indicated by 
the shaded region.
The lowest six moments of this average valence $u_v-d_v$ distribution are
calculated to be:
\begin{eqnarray}
  \label{avgmoments}
  \langle x^0\rangle_{u_v-d_v}^{\rm avg}=1\ ,\quad
  \langle x^1\rangle_{u_v-d_v}^{\rm avg}=0.163\ ,\quad 
  \langle x^2\rangle_{u_v-d_v}^{\rm avg}=0.0543\ ,\nonumber \\
  \langle x^3\rangle_{u_v-d_v}^{\rm avg}=0.0229\ ,\quad
  \langle x^4\rangle_{u_v-d_v}^{\rm avg}=0.0111\ ,\quad
  \langle x^5\rangle_{u_v-d_v}^{\rm avg}=0.00599\ . \nonumber
\end{eqnarray}
We have investigated fits to these moments using the following seven
fitting functions:
(i) the full form of Eq.~(\ref{momentfit}) with unconstrained parameters;
(ii) a simplified form with $\gamma=0$;
(iii) a form with $\epsilon=0$;
(iv) both $\epsilon=0$ and $\gamma=0$;
(v) with the constraint $\epsilon=\epsilon_\Delta$;
(vi) $\gamma=\gamma_\Delta$; and finally
(vii) $\epsilon=\epsilon_\Delta$ and $\gamma=\gamma_\Delta$.
Fits (iii) and (iv) are unable to reproduce the small $x$ behavior of the
distribution and are therefore discarded. 
The parameters determined by $\chi^2$ fits with each of the remaining
forms give rise to similar distributions.
We use the differences in the values of the parameters found from the 
various fits as an indication of the systematic error in the 
reconstruction procedure.
In particular, for the parameter $b$ (which is of special interest
because of its relation to Regge phenomenology--see Section~IV below),
we find the best fit value $b = 0.4 \pm 0.2$.

Figure~\ref{reconstruct} shows the reconstructed $x (u_v(x)-d_v(x))$
distribution (short--dashed line) which is generated by the fit form (i)
using all six moments above. 
With just the lowest 4 moments, however (as are available from lattice
data), this form does not produce an accurate fit, as one is attempting
to constrain more parameters than the number of data points permit%
  \footnote{Here we count the zeroth moment as a fit point with zero error, and include
  the overall normalization parameter $a$ in our parameter counting.}.
This is illustrated by the dotted line in the Fig.~\ref{reconstruct}.
In contrast, with fit (vii), which has only 3 free parameters, one can
obtain a fairly reliable reconstruction even when fitting to only the
$n=0,1,2$ moments, as indicated by the solid curve in
Figure~\ref{reconstruct}.

As well as understanding the systematic uncertainty associated with the
functional form of the distribution to be reconstructed, one must also
quantify the error arising from the use of only a fixed number of moments.
When the number of moments used with fitting form (vii) is increased to
4, 5 or 6 the fit improves by only a few percent. 
With four moments, fits (ii), (v) and (vi) also give acceptable results,
similar to those shown for fit (vii).
For simplicity, however, we will concentrate on fit form (vii) hereafter.

The overall agreement between the original and reconstructed distributions
is clearly excellent, and comparable to the difference between different
parameterizations (shaded region in Fig.~\ref{reconstruct}).
Having confidence that a reliable reconstruction is possible from just the
lowest few moments, we next ask what do extrapolations of lattice data on
the $n=0,1,2,3$ moments tell us about the $x$ dependence of the distribution
$u_v(x)-d_v(x)$?

%%%%%%%%%%%%%%%%%%%%%%%%%%%%%%%%%%%%%%%%%%%%%%%%%%%%%%%%%%%%%%%%%%%%%%%%%
\section{Quark Distributions from Lattice Moments}

A crucial question for relating lattice calculations to phenomenology
is the reliability of the extrapolation from the region of large quark
masses ($m_q$) to the physical value.
Typically, lattice data have been analyzed assuming a linear dependence
on the quark mass.
However, as pointed out in Ref.\cite{DMNRT}, if one is to extrapolate the
lattice moments into a region where pion loops play an important role,
it is vital to use an extrapolation formula which has the correct chiral
behavior.
In this section we first motivate an extrapolation formula which ensures
both the correct chiral and heavy quark limits.
Following this, we use the values of the moments at physical quark masses
extracted from the lattice data to reconstruct the $x$ dependence of the
nonsinglet distribution.

% .......................................................................
\subsection{Constraints from the Chiral and Heavy Quark Limits}

The spontaneous breaking of the chiral SU$_L$(N$_f$)$\times$SU$_R$(N$_f$)
symmetry of QCD generates the nearly massless Goldstone bosons (pions),
whose importance in hadron structure is well documented.
At small pion masses, hadronic observables can be systematically expanded
in a series in $m_\pi$ \cite{CHIPT}.
The expansion coefficients are generally free parameters, determined from
phenomenology or obtained from low energy models.
One of the unique consequences of pion loops, however, is the appearance
of so-called chiral logs.
{}From the Gell-Mann--Oakes--Renner relation one finds that
$m_\pi^2 \sim m_q$ at small $m_\pi$, so that these terms are nonanalytic
functions of the quark mass.
Because the nonanalytic terms arise from the infra-red behavior of the
chiral loops, they are generally model independent.
For some observables these nonanalytic terms give rise to large deviations
from linearity near the chiral limit \cite{OBS}.

\begin{figure}
\begin{center}
  \epsfysize=16truecm \leavevmode
  \epsfbox{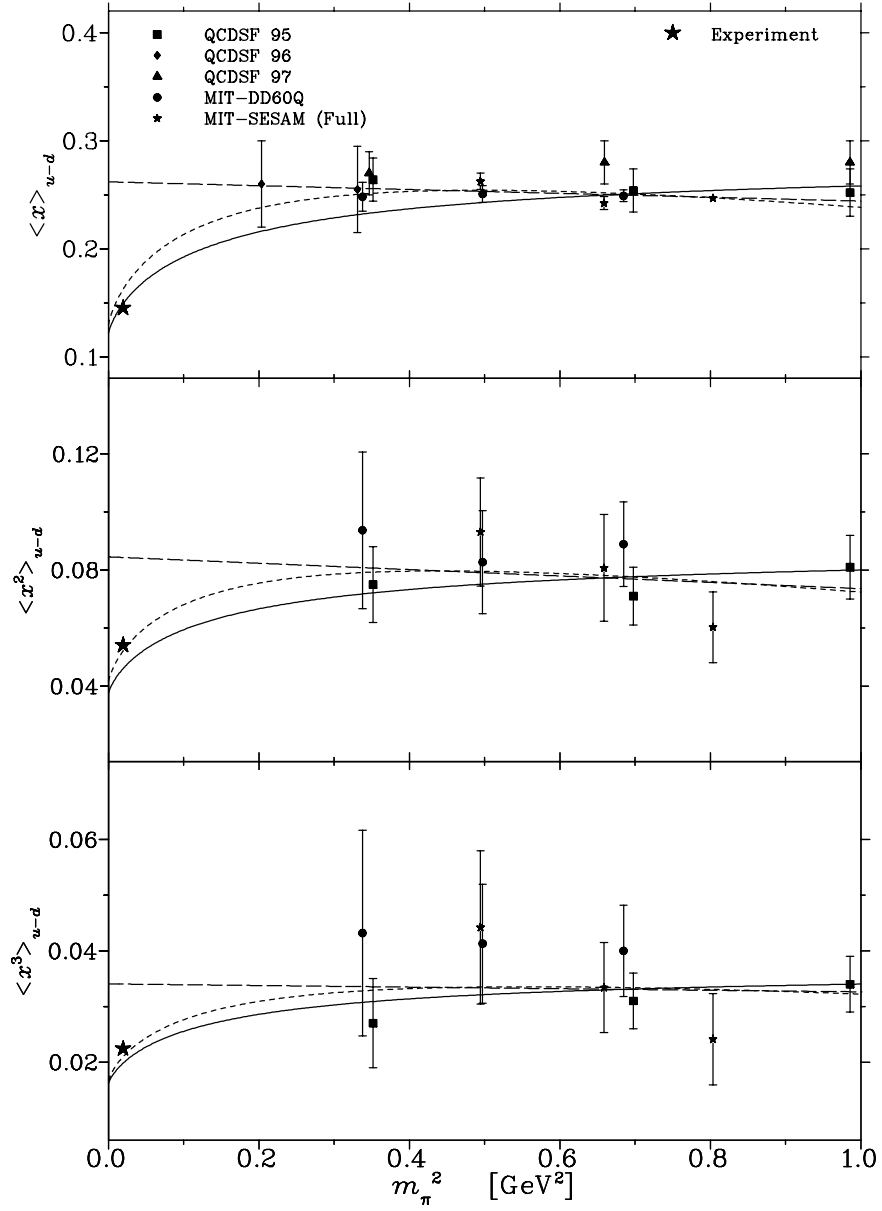}\vspace*{1cm}
\caption{Moments of the $u-d$ distribution.
        The lattice data \protect\cite{QCDSF,MIT} are fitted using a
        linear function (long-dashed), and with the improved chiral
        extrapolation in Eq.~(\protect\ref{xtrap}) (solid).
        Also shown is the fit from Ref.\protect\cite{DMNRT} 
        (short-dashed) which does
        not impose the heavy quark limit, Eq.(\protect\ref{HQlimit}).}
\label{fitcomparison}
\end{center}
\end{figure}

The leading order (in $m_\pi$) nonanalytic term in the expansion of the
moments of PDFs was shown in Ref.\cite{TMS} to have the generic behavior
$m_\pi^2 \log m_\pi$.
In Ref.\cite{DMNRT} a low order chiral expansion for the moments of the
nonsinglet $u-d$ distribution was developed which incorporated the leading
nonanalytic (LNA) behavior of the moments as a function of $m_q$.
Here we extend this in order to connect also with the heavy quark limit,
in which the valence quark distributions become $\delta$-functions
centered at $x=1/3$:
\begin{equation}
  u(x)-d(x)\stackrel{m_q\rightarrow\infty}{\longrightarrow}
  \delta(x-\sfrac{1}{3})\ .
\end{equation}
The corresponding moments in the heavy quark limit are therefore given by:
\begin{equation}
\label{HQlimit}
  \langle x^n\rangle_{u-d}
  \stackrel{m_q\rightarrow\infty}{\longrightarrow}
  \frac{1}{3^n}\ .
\end{equation}
It is straightforward to incorporate this constraint into the
extrapolation formula, and its inclusion makes no significant difference
to the central results of Ref.\cite{DMNRT}, as illustrated in
Fig.~\ref{fitcomparison}.
For completeness, we give the extrapolation formula which explicitly
satisfies both the heavy quark and chiral limits:
\begin{equation}
\label{xtrap}
\langle x^n\rangle_{u-d}\
=\ a_n \left( 1 + c_{\rm LNA}m_\pi^2\log\frac{m_\pi^2}{m_\pi^2+\mu^2}
       \right)\
+\ b_n\frac{m_\pi^2}{m_\pi^2+m_{b,n}^2}\ ,
\end{equation}
where the coefficient $c_{\rm LNA} = -(1 + 3 g_A^2)/(4\pi f_\pi)^2$
\cite{CHPT}, and $b_n$ is a constant constrained by Eq.(\ref{HQlimit}):
\begin{equation}
  b_n = \frac{1}{3^n} - a_n \left( 1 - \mu^2 c_{\rm LNA} \right)\ .
\end{equation}
In this extrapolation scheme there are therefore three free parameters
for each moment: $a_n,\,m_{b,n}$, and $\mu$.
We find that the dependence of the mass parameter $m_{b,n}$ on 
$n$ is quite small, and therefore fix $m_{b,n}=m_b$ for all $n$.
As the $b_n$ term in Eq.~(\ref{xtrap}) is included in order to provide a
linear dependence on $m_{\pi}^{2}$ in the region where lattice data exist,
$m_b$ should be large enough such that this is indeed the case.
For $m_b > 4$~GeV the fits to the moments are essentially independent
of $m_b$; we therefore fix $m_b=5$~GeV.

The mass $\mu$ essentially determines the scale above which pion loops no
longer yield rapid variation.
Analyses of a wide range of observables such as masses, magnetic moments
and charge radii \cite{OBS} as well as structure function moments suggest 
a common scale, $\mu \sim 0.5$~GeV, which simply reflects the scale at 
which the Compton wavelength of the pion becomes comparable to the size of 
the hadron (without its pion cloud).
We use the value $\mu = 550$~MeV found in our previous fits to the lattice
moments of the $u-d$ distributions in Ref.\cite{DMNRT}. 
Consequently the extrapolation for each moment is essentially
parameterized by only one variable.
Nevertheless, the best fit using Eq.(\ref{HQlimit}) is as good as that
found in our earlier work \cite{DMNRT} (which did not impose the additional constraint of the heavy
quark limit), as illustrated in Fig.~\ref{fitcomparison} --- the latter is 
given by the short--dashed curve and the former by the solid curve.
For details of the fitting procedure and the error analysis we refer to
Ref.\cite{DMNRT}.

Having motivated the functional form of the extrapolation formula to
connect with both the chiral and heavy quark limits, we now use the
extracted moments to calculate the $x$ dependence of the nonsinglet
valence $u-d$ distribution from the lattice data.

% .......................................................................
\subsection{Results}

Before extracting information on the $x$ dependence of a particular PDF 
from the lattice moments, one must note that the crossing symmetry 
properties of the spin-averaged quark distributions (see
Eq.~(\ref{moments})) mean that the moments of $u-d$ calculated on the
lattice correspond to the {\em valence} $u_v-d_v$ distribution only for
{\em even} $n$.
The {\em odd} $n$ moments, on the other hand, correspond to the
combination $u + \bar u - d - \bar d = u_v - d_v - 2 (\bar d - \bar u)$.  
The difference between the two represents the famous violation of the 
Gottfried sum rule, when $\bar d \not= \bar u$ \cite{NMC,E866}.
Some care must be taken, therefore, if one attempts to use information on 
both even and odd moments to extract a valence quark distribution.
To minimize possible contamination of the extracted valence distribution 
arising from the flavor asymmetry of the sea, in the present analysis we 
subtract the values of the phenomenological moments of the $\bar d-\bar u$ 
difference, as given by global parameterizations of data 
\cite{MRS,CTEQ,GRV}, from the calculated odd moments.

Phenomenologically, the difference $\bar d(x) - \bar u(x)$ is found to be 
concentrated at small $x$\ ($\bar d - \bar u \sim (1-x)^{14}$ at
$Q^2 \sim 50$~GeV$^2$ from the Fermilab Drell-Yan data \cite{E866}),
and will therefore become progressively less important for larger $n$
compared with the valence distribution.
The largest effect is for the $n=1$ moment and results in a correction of
$< 10\%$ of the value of $\langle x \rangle_{u-d}$ in
Eq.~(\ref{avgmoments}), if one adds the phenomenological value
$2\int_0^1 x (\bar d(x) - \bar u(x))dx=0.017(2)$ \cite{MRS,CTEQ,GRV} to
the extrapolated first moment.
For the $n=3$ moment, the corresponding correction is a fraction of a
percent.
In principle, given sufficiently many moments, one can fit {\em both} the
total $u-d$ and valence $u_v-d_v$ distributions separately from the $n$ 
even and $n$ odd moments, respectively --- however, given the small number 
of moments currently available, the extraction of the valence distribution 
is the most feasible solution.

\begin{figure}
\begin{center}
  \epsfysize=9truecm \leavevmode
  \epsfbox{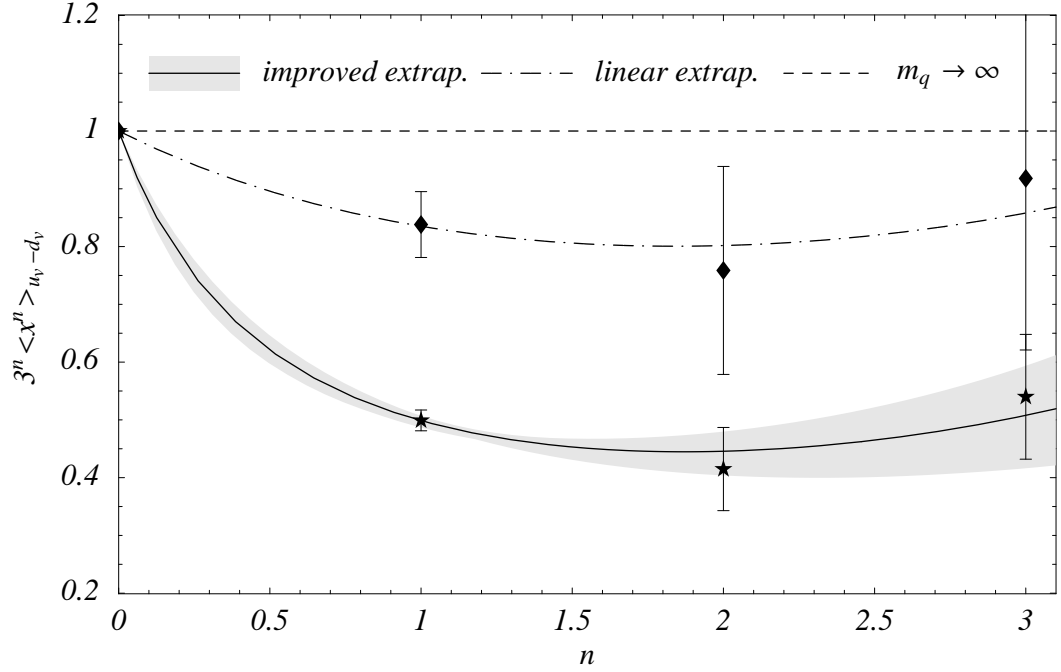}\vspace*{1cm}
  \caption{The lowest four moments of the valence $u_v-d_v$ distribution
        (scaled by $3^n$) at the physical quark mass, extracted from the
        fit to the lattice data using a linear extrapolation (diamonds)
        and the improved chiral extrapolation, Eq.~(\protect\ref{xtrap})
        (stars).
        The solid and dot-dashed lines are $\chi^2$ fits to the improved
        and linearly extrapolated moments respectively using
        Eq.~(\protect\ref{momentfit}), with $\epsilon$ and $\gamma$
        constrained to their average values.
        The shaded region represents a 1 standard deviation of the fit
        parameters about the optimal values for the improved
        extrapolation.
        The short-dashed line represents the heavy quark limits of the moments.}
\label{lvcmom}
\end{center}
\end{figure}

Table~I lists the resulting extrapolated values for the lowest three
nontrivial moments of the {\em total} $u-d$ distribution at the
physical quark mass for both a linear extrapolation and the chirally
symmetric extrapolation of Eq.~(\ref{xtrap}).
Also listed are the corresponding values of the valence fit parameters
$a$, $b$ and $c$ from $\chi^2$ fits.
As discussed in Section~II, because there are only four data points with
which to fix parameters, we use Eq.~(\ref{momentfit}) with the fitting
form (vii), namely fixing $\epsilon=\epsilon_\Delta$ and
$\gamma=\gamma_\Delta$.
With these parameters, the resulting fits to the moments of the
{\em valence} $u_v-d_v$ distribution are displayed in Fig.~\ref{lvcmom}
for both the linear and improved, chirally symmetric extrapolations.
For clarity we plot $3^n$ times the moments, so that the horizontal line 
at unity represents the heavy quark limit of the moments.
The lightly shaded region around the moments from the improved chiral
extrapolation corresponds to a 1 standard deviation variation of the fit
parameters from their optimal values.

\begin{figure}
\begin{center}
  \epsfysize=9truecm \leavevmode 
  \epsfbox{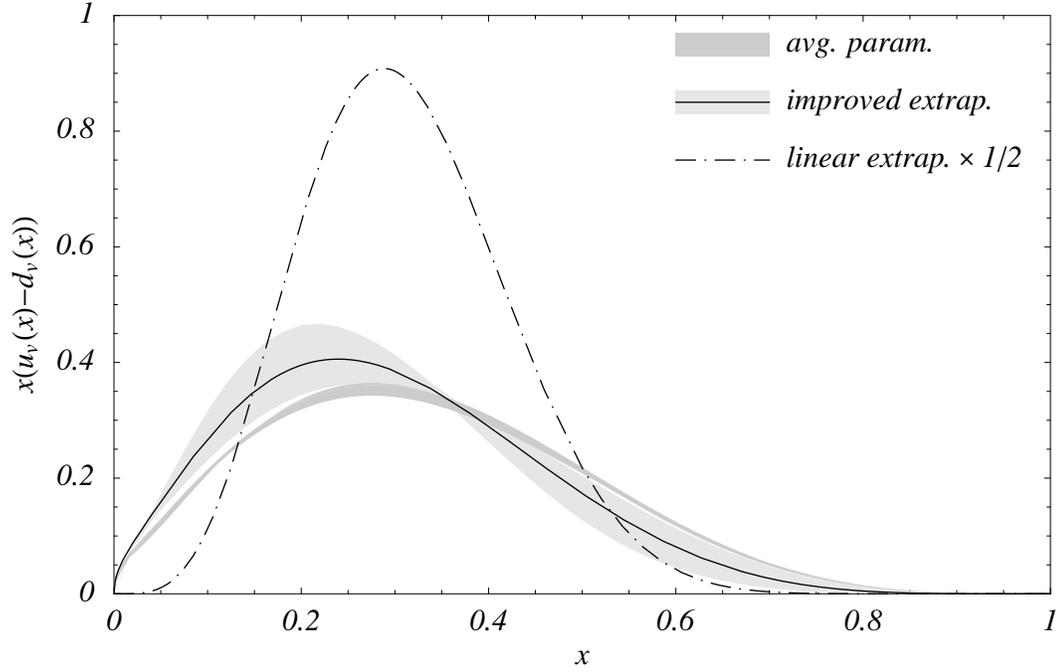}\vspace*{1cm}
  \caption{The physical valence $x(u_v(x)-d_v(x))$ distribution,
        extracted using the improved chiral extrapolation of the lattice 
        moments (solid), and a linear extrapolation, scaled by a factor
        1/2 (dot--dashed).
        The lighter shaded region indicates a $1 \sigma$ variation of the
        fit parameters about the optimal values for the improved
        extrapolation, while the dark shaded region represents the spread
        between global parameterizations \protect\cite{MRS,CTEQ,GRV}.}
\label{lvcpdf}
\end{center}
\end{figure}

The corresponding distributions $x (u_v(x)-d_v(x))$ are displayed in
Fig.~\ref{lvcpdf}.
Once again, the lightly shaded region represents a $1 \sigma$ deviation
from the central values, while the darker band illustrates the spread 
between the global parameterizations from Refs.\cite{MRS,CTEQ,GRV}.
A comparison of the distribution reconstructed from the lattice data using 
the improved chiral extrapolation with the phenomenological distributions 
shows reasonably good agreement.
On the other hand, the linear extrapolation gives a distribution (scaled 
by a factor $1/2$ in the figure) which is much smaller than the 
phenomenological distributions at small $x$, and consequently has a 
significantly higher peak in the intermediate $x$ region, centered at 
$x \sim 1/3$ -- reminiscent of a heavy, constituent quark--like 
distribution.

\begin{figure}
\begin{center}
  \epsfysize=9truecm \leavevmode
  \epsfbox{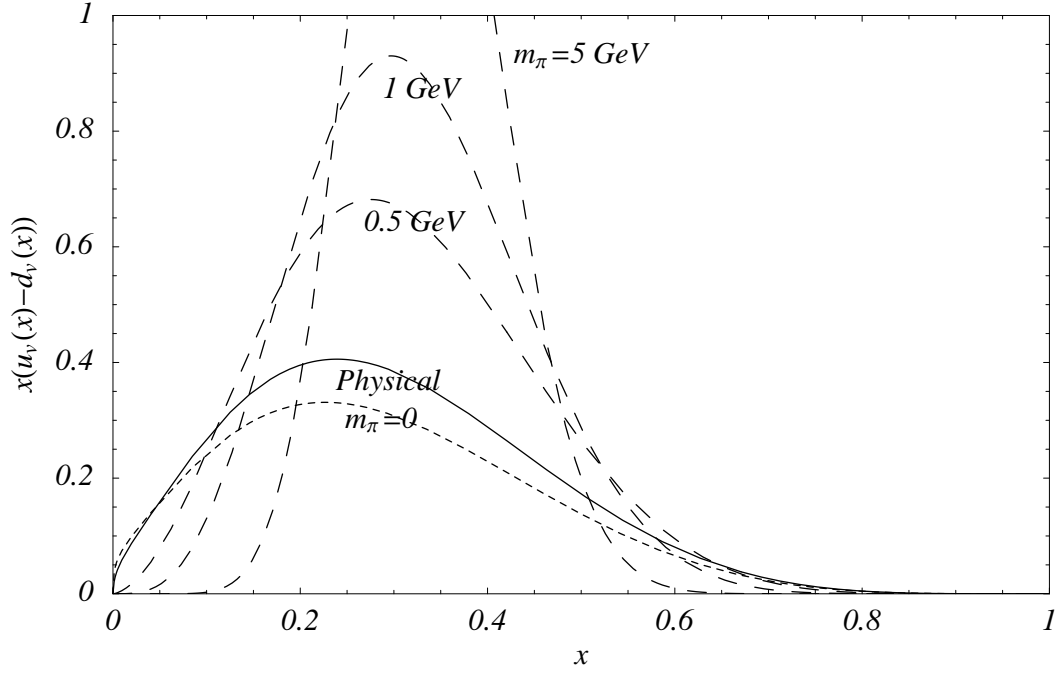}\vspace*{1cm}
  \caption{The nonsinglet valence distribution $x(u_v(x)-d_v(x))$
        extracted from the improved extrapolation formula,
        Eq.~(\protect\ref{xtrap}), for various pion masses: $m_\pi=0$
        (short-dashed), $m_\pi=0.139$~GeV (solid), $m_{\pi}=0.5$, 1
        and 5~GeV
        (long-dashed).  The fit parameters are tabulated in Table~II.}
\label{pdfvsmass}
\end{center}
\end{figure}

The improved extrapolation formula (\ref{xtrap}) enables one to study not
only the quark distribution at the physical quark mass, but also the
dependence of the distribution on the quark (or equivalently, pion) mass.
In particular, it allows one to trace how the $x$ dependence changes as
one goes from the chiral limit to the heavy quark limit, where the 
distribution approaches a $\delta$-function.
Table~\ref{massdep} lists the extrapolated values of the total moments at
various pion masses.
Fitting the corresponding valence moments with the functional form (vii) 
from Section~II, this table also shows how the resulting fit parameters 
change as the pion mass is increased from $m_\pi=0$ to $m_\pi=5$~GeV.
Notice that the fit parameters $b$ and $c$ become larger with increasing 
$m_\pi$, as the $x$ dependence of the valence distribution becomes less 
singular at small $x$.
This is dramatically illustrated in Fig.~\ref{pdfvsmass}, where the 
corresponding valence $x (u_v-d_v)$ distribution is shown in the chiral 
limit ($m_\pi=0$), at the physical mass ($m_\pi = 0.139$~GeV), and at 
$m_\pi = 0.5$, 1 and 5~GeV.
As $m_\pi$ increases, the distribution becomes visibly more sharply
peaked, with the peak moving towards the limiting value of $x=1/3$.

Quite remarkably, the distribution at $m_\pi=5$~GeV resembles the
distribution in Fig.~\ref{lvcpdf} (dot--dashed curve) extracted from the
linearly extrapolated moments.
Since the linearly extrapolated moments are not constrained by the
expected behavior in the heavy quark limit, it is noteworthy that a
constituent quark--like distribution, peaking near $x \sim 1/3$, 
nevertheless arises if the moments are flat as a function of $m_\pi$,
as one would expect within a constituent quark model.
This in fact provides further {\em a posteriori} justification for the
choice made in Eq.(\ref{xtrap}) to build in the correct heavy quark limit
(as well as the correct chiral behavior).

%%%%%%%%%%%%%%%%%%%%%%%%%%%%%%%%%%%%%%%%%%%%%%%%%%%%%%%%%%%%%%%%%%%%%%%%%
\section{Regge Behavior and the $J^{PC}=1^{--}$ Trajectory}

In the previous section we extracted the exponent $b$ which governs the 
small-$x$ behavior of the nonsinglet valence quark distribution $u_v-d_v$.  
According to Regge theory \cite{COLLINS}, this exponent is related to the 
slope of the isovector $J^{PC}=1^{--}$ trajectory, corresponding to the 
exchange of $\rho$ ($1^{--}$), $a_2$ ($2^{++}$), $\rho_3$ ($3^{--}$) and so on.
If one makes the reasonable hypothesis that Regge theory also holds for
quarks with masses in the range $m_q \sim 30$--190~MeV used in lattice
simulations, one can estimate the masses of mesons on this trajectory
from fits to the valence quark moments versus $m_q$.

{}From Regge theory the elastic scattering amplitude at high center of
mass energy, $s$, behaves as:
\begin{equation}
{\cal A}(s,t) \sim s^{\alpha(t)}\ ,
\end{equation}
where $\alpha(t)$ is a function describing the meson trajectory:
\begin{equation}
  \alpha(t) = \alpha_0 + \alpha^\prime\ t\ ,
\end{equation}
with $\alpha^\prime$ the slope and $\alpha_0$ the intercept of the
trajectory.
The optical theorem then relates the imaginary part of the scattering 
amplitude at forward angles, ${\cal A}(s,t=0)$, to the total cross 
section, $\sigma$:
\begin{equation}
\sigma\ \sim\ {{\rm Im} {\cal A} \over s}\ \sim\ s^{\alpha_0 - 1}\ .
\end{equation}
At large $s$ the approximate energy independence of total cross
sections is usually attributed to the exchange of the Pomeron
(${\rm I\!P}$) trajectory, which has intercept
$\alpha_0^{\rm I\!P} \approx 1$.
On the other hand, the isovector cross sections (corresponding to nonsinglet distributions) are dominated by the 
exchange of a trajectory with $\bar q q$ quantum numbers corresponding to 
the $\rho$ meson and its excitations, which has intercept
$\alpha_0^{\rho} \approx 0.5$.

In deep-inelastic scattering at large $s \gg Q^2$, corresponding to
$x = Q^2/2p\cdot q \sim Q^2/s \ll 1$, the total virtual photon--nucleon
cross section is dominated by the singlet quark and gluon distributions, 
and behaves as
$\sigma^{\gamma^* N} \sim F_2^N \sim x^{1 - \alpha_0^{\rm I\!P}}$.
The nonsinglet distribution, $q_{\rm NS}$, on the other hand, is governed 
by the exchange of the $J^{PC}=1^{--}$ trajectory,
$x q_{\rm NS} \sim x^{1 - \alpha_0^{\rho}} \sim x^{1/2}$
(recall that the best fit to the average $u_v-d_v$ difference in
Eq.~(\ref{avgdiff}) gave an exponent $b=0.476$).
Identifying $\alpha_0^{\rho} \approx 1-b$ from the previous section as
a function of quark mass, one can therefore extract information on the
$m_q$ dependence of the Regge intercept from the $m_q$ dependence of
the nonsinglet quark distribution.

Having determined the intercept $b \approx 1-\alpha_0^{\rho}$ from the PDF 
fits, one needs to determine the slope of the trajectory as a function of 
$m_q$.
In the infinite mass limit, orbital excitations of mesons become 
energetically degenerate with the $L=0$ state.
Within the Regge picture, this is possible only if the Regge intercept
$\alpha_0^{\rho} \stackrel{m_q\to\infty}{\longrightarrow} -\infty$, or
$b \stackrel{m_q\to\infty}{\longrightarrow} \infty$, which is consistent
with the valence distribution approaching a $\delta$-function.
One expects, therefore, that the slope $\alpha^\prime$ should increase
as $m_q$ increases from its physical value.
The behavior
$\alpha_0^\rho \stackrel{m_q\to\infty}{\longrightarrow} -\infty$ is
indeed what is seen from the $m_\pi$ dependence of $b$ in Table~II.
The slope at larger, but finite, values of $m_q$ is most accurately
determined from the mass of the $\rho$ meson, which has considerably
smaller errors than the masses of higher lying mesons on the same
trajectory.
Using the lattice data for the $\rho$ meson and the values of 
$\alpha_0^{\rho}(m_q)$ generated from the fits in Section~III, one can 
then make predictions for the behavior of the masses of the orbital 
excitations as a function of $m_q$.

\begin{figure}
\begin{center}
  \epsfysize=9truecm \leavevmode
  \epsfbox{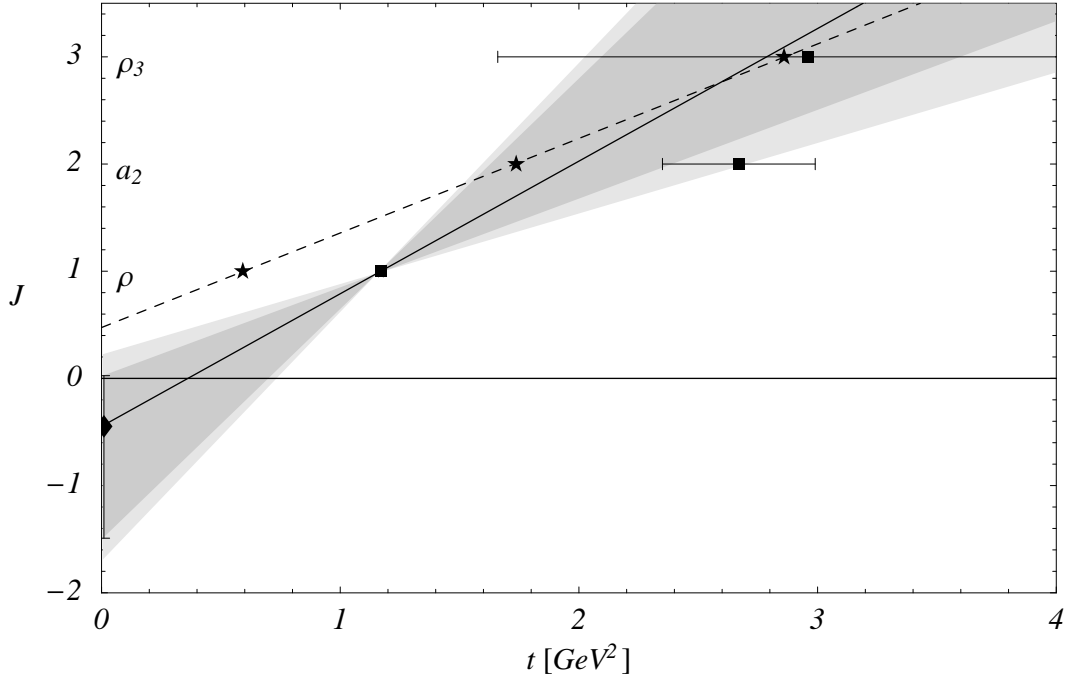}\vspace*{1cm}
  \caption{Regge plot of the spin, $J=\alpha_0^\rho$, versus
        $t=({\rm mass})^2$ of mesons on the $J^{PC}=1^{--}$ 
        trajectory, at the physical pion mass (dashed) and at 
        $m_{\pi}=785$~MeV (solid), with the point at $t=0$ obtained from 
        the best fit to $b$ (see Table~II).
        The darker shaded region, which includes the error on $b$,
        illustrates the statistical error in the extrapolation of the 
        lattice moments, while the lighter shader region indicates the 
        systematic error in the fitting procedure.
        The physical masses of the $\rho$, $a_2$ and $\rho_3$ mesons are
        indicated by stars, while the boxes represent lattice
        masses at $m_\pi=0.785$~GeV from the UKQCD Collaboration
        \protect\cite{ORBIT}.}
\label{Reggetraj}
\end{center}
\end{figure}

In Fig.~\ref{Reggetraj} we show the predicted $1^{--}$ Regge trajectory at 
$m_\pi = 0.785$~GeV (solid line), compared with the trajectory at the 
physical light quark mass (dashed).
A fit through the central values of the parameter $b$ and the $\rho$ mass 
at $m_\pi = 0.785$~GeV (as calculated in Ref.~\cite{ORBIT}) yields a slope 
which is somewhat larger than that of the trajectory at the physical quark
mass, which is consistent with the expected trend towards the heavy quark
limit.
The darker shaded region represents the statistical error in the 
extrapolation of the lattice moments, obtained from fitting the upper 
and lower bounds of the original lattice data.
The lighter region indicates an estimate of the systematic error 
associated with the fitting procedure as discussed in Section~II.

Although lattice data for the masses of orbital excitations are scarce,
there have been some pioneering calculations of masses of the $a_2$
($J^{PC}=2^{++}$) and $\rho_3$ ($J^{PC}=3^{--}$) mesons by the UKQCD
Collaboration \cite{ORBIT}, indicated by the filled boxes in
Fig.~\ref{Reggetraj}.
Comparing with the predictions from the PDF analysis, the lattice data are 
in reasonable agreement: the calculated $\rho_3$ meson mass lies within 
the predicted band, albeit within large errors, while the $a_2$ mass lies 
on the edge of the predicted range.
Needless to say, further exploration of the masses of excited mesons 
within lattice QCD would be very helpful in testing these predictions.

%%%%%%%%%%%%%%%%%%%%%%%%%%%%%%%%%%%%%%%%%%%%%%%%%%%%%%%%%%%%%%%%%%%%%%%%%
\section{Conclusion}

In this paper we have extracted the $x$ dependence of the unpolarized
valence quark distribution $x (u_v(x)-d_v(x))$ from lattice QCD data
on the moments of the $u-d$ distribution.
After establishing that the basic features of the $x$ distributions can be 
reconstructed from just the lowest four moments, we analyze the lattice 
data using an extrapolation formula which embodies the correct behavior in 
both the chiral and heavy quark limits.

The importance of ensuring the correct chiral behavior is illustrated by 
comparing the $x$ distributions obtained by extrapolating the lattice data 
using a linear and an improved chiral extrapolation.
While the improved extrapolation gives an $x$ distribution which is in
quite good agreement with the phenomenological parameterizations, the
linearly extrapolated data give rise to distributions with the wrong
small-$x$ behavior, which translates into a much more pronounced peak at 
$x \sim 1/3$, reminiscent of a heavy, constituent quark--like
distribution.
The approach to the heavy quark limit is explicitly mapped out by studying 
the dependence of the $x$ distributions on the pion mass, from the chiral 
limit to pion masses of several GeV.

Our analysis suggests an intriguing connection between the small-$x$
behavior of the valence distributions and the $m_q$ dependence of meson
masses on the $J^{PC}=1^{--}$ trajectory.
While the existing lattice data are in qualitative agreement with the
predictions from the PDF analysis, future results on the masses of excited
meson states at varying quark mass will enable the accuracy of this
relation to be tested more thoroughly.
Plans by a number of lattice groups to study the spectrum of excited 
hadrons should provide valuable insights into this problem.

%%%%%%%%%%%%%%%%%%%%%%%%%%%%%%%%%%%%%%%%%%%%%%%%%%%%%%%%%%%%%%%%%%%%%%%%%
\acknowledgements

We would like to thank D.B.~Leinweber and C.~Michael for helpful discussions
and communications.
This work was supported by the Australian Research Council,
and the U.S. Department of Energy contract \mbox{DE-AC05-84ER40150},
under which the Southeastern Universities Research Association (SURA)
operates the Thomas Jefferson National Accelerator Facility
(Jefferson Lab).

%%%%%%%%%%%%%%%%%%%%%%%%%%%%%%%%%%%%%%%%%%%%%%%%%%%%%%%%%%%%%%%%%%%%%%%%%

%%%%%%%%%%%%%%%%%%%%%%%%%%%%%%%%%%%%%%%%%%%%%%%%%%%%%%%%%%%%%%%%%%%%%%%%%
% TABLE
%%%%%%%%%%%%%%%%%%%%%%%%%%%%%%%%%%%%%%%%%%%%%%%%%%%%%%%%%%%%%%%%%%%%%%%%%

\begin{table}
\begin{tabular}{||l||l|l|l||l|l|l||}
Extrapolation & $\langle x^1\rangle$ & $\langle x^2\rangle$ 
& $\langle x^3\rangle$ & $a$ & $b$ & $c$ \\
\hline
Linear     & 0.262 & 0.0843 & 0.0340 & 658   & 3.51  & 10.5 \\
Improved   & 0.149 & 0.0461 & 0.0200 & 0.439 & 0.351 & 3.77 \\
\end{tabular}
\vspace*{0.5cm}
\caption{\label{linvschi}
        Moments of the {\em total} $u-d$ distribution obtained from
        a linear extrapolation and the improved chirally symmetric
        extrapolation, Eq.~(\ref{xtrap}), to the physical quark mass.
        Also shown are the best fit parameters $a$, $b$ and $c$ for
        the {\em valence}
        $u_v(x)-d_v(x)$ distribution, using Eq.~(\ref{momentfit}) with
        $\epsilon=\epsilon_\Delta$ and $\gamma=\gamma_\Delta$ constrained 
        to the average values (see text).}
\end{table}

\begin{table}
\begin{tabular}{||c||c|c|c||c|c|c||}
$m_\pi$ (GeV)& $\langle x^1\rangle$ & $\langle x^2\rangle$ 
& $\langle x^3\rangle$ & $a$ & $b$ & $c$ \\
\hline
0     & 0.121 & 0.0374 & 0.0161 & 0.282  & 0.245 & 3.55 \\
$m_\pi^{\rm phys}$
      & 0.149 & 0.0461 & 0.0200 & 0.439  & 0.351 & 3.77 \\
0.5   & 0.223 & 0.0689 & 0.0296 & 2.39   & 0.957 & 4.96 \\
0.785 & 0.248 & 0.0769 & 0.0328 & 7.30   & 1.45  & 5.91 \\
1     & 0.258 & 0.0801 & 0.0341 & 13.8   & 1.74  & 6.48 \\
5     & 0.303 & 0.0978 & 0.0367 & 125000 & 6.20  & 15.4 \\
\end{tabular}
\vspace*{0.5cm}
\caption{\label{massdep}
        Moments of the {\em total} $u-d$ distribution obtained from
        the improved chiral extrapolation formula, Eq.~(\ref{xtrap}),
        at various pion masses; and best fit parameters for the
        corresponding {\em valence} $u_v(x)-d_v(x)$ distribution
        using Eq.~(\ref{momentfit}) with $\epsilon=\epsilon_\Delta$ 
        and $\gamma=\gamma_\Delta$ constrained to the average values.}
\end{table}

%%%%%%%%%%%%%%%%%%%%%%%%%%%%%%%%%%%%%%%%%%%%%%%%%%%%%%%%%%%%%%%%%%%%%%%%%
% FIGURES
%%%%%%%%%%%%%%%%%%%%%%%%%%%%%%%%%%%%%%%%%%%%%%%%%%%%%%%%%%%%%%%%%%%%%%%%%

\newpage

\newpage

\newpage

\newpage

\newpage

\end{document}